\documentclass[conference]{IEEEtran}

\usepackage[T1]{fontenc}
\usepackage{graphicx}
\usepackage{cite}
\usepackage{caption}
\usepackage{subcaption}
\usepackage{epstopdf}

\usepackage{overpic}
\usepackage{amssymb}
\usepackage{amsmath}
\usepackage{amsthm}
\usepackage{array}
\usepackage{color}
\usepackage{url}

\IEEEoverridecommandlockouts

\theoremstyle{plain}
\newtheorem{theorem}{Theorem}

\def\sinc{\mathrm{sinc}}

\def\CN{\mathcal{N}_{\mathbb{C}}} 
\def\imagunit{\mathsf{j}} 

\begin{document}

\title{A Primer on Near-Field Beamforming for Arrays and Reconfigurable Intelligent Surfaces}

\author{
\IEEEauthorblockN{Emil Bj{\"o}rnson\IEEEauthorrefmark{1}\IEEEauthorrefmark{3}, \"Ozlem Tu\u{g}fe Demir\IEEEauthorrefmark{1}, 
Luca Sanguinetti\IEEEauthorrefmark{2}
\thanks{This work was supported by the  FFL18-0277 grant from the SSF.}}
\IEEEauthorblockA{\IEEEauthorrefmark{1}Department of Computer Science, KTH Royal Institute of Technology, Stockholm, Sweden (\{emilbjo, ozlemtd\}@kth.se)}
\IEEEauthorblockA{\IEEEauthorrefmark{2}Dipartimento di Ingegneria dell'Informazione, University of Pisa, 56122 Pisa, Italy (luca.sanguinetti@unipi.it)}
\IEEEauthorblockA{\IEEEauthorrefmark{3}Department of Electrical Engineering (ISY), Link\"{o}ping University, Link\"{o}ping, Sweden}
}

\maketitle

\begin{abstract} 
Wireless communication systems have almost exclusively operated in the far-field of antennas and antenna arrays, which is conventionally characterized by having propagation distances beyond the Fraunhofer distance. This is natural since the Fraunhofer distance is normally only a few wavelengths. With the advent of active arrays and passive reconfigurable intelligent surfaces (RIS) that are physically large, it is plausible that the transmitter or receiver is located  in between the Fraunhofer distance of the individual array/surface elements and the Fraunhofer distance of the entire array. An RIS then can be configured to reflect the incident waveform towards a point in the radiative near-field of the surface, resulting in a beam with finite depth, or as a conventional angular beam with infinity focus, which only results in amplification in the far-field.
To understand when these different options are viable, an accurate characterization of the near-field behaviors is necessary. In this paper, we revisit the motivation and approximations behind the Fraunhofer distance and show that it is not the right metric for determining when near-field focusing is possible.
We obtain the distance range where finite-depth beamforming is possible and the distance where the beamforming gain tapers off.
\end{abstract}


\vspace{-2mm}

\IEEEpeerreviewmaketitle

\section{Introduction}

\vspace{-1mm}

\enlargethispage{4mm}

When an antenna radiates a wireless signal in free-space, the field distribution is uniquely determined by Maxwell's equations but the wavefront anyway appears to have a different shape depending on the observation distance. Traditionally, three regions were distinguished \cite{Sherman1962a}: the near-field, the Fresnel region, and the far-field.
These regions are defined from the transmitter perspective but can be equivalently viewed from the receiver due to reciprocity \cite{Selvan2017a}. We will consider both perspectives.
If the receive antenna is in the (radiative) \emph{near-field} of the transmitter, the propagation distances are so short that there are noticeable amplitude variations over the receiver aperture \cite{Friedlander2019a}.
These variations are negligible in the \emph{Fresnel region} but remain to be substantial compared to the wavelength, thus causing noticeable phase variations over the receiver aperture. Finally, both the amplitude and phase variations are negligible in the \emph{far-field}, where the amplitude depends only on the propagation distance to the center of the receiver and the phase variations only depend on the incident angle. 
Contemporary wireless communication systems operate in the far-field, but this might change in the future when the antenna apertures increase \cite{Bjornson2019d} and the wavelength shrinks \cite{Rappaport2019a}.

The near-field and Fresnel regions of aperture antennas have been thoroughly studied in the antennas and propagation literature  \cite{Polk1956a,Kay1960a,Sherman1962a,Hansen1985a,Nepa2017a}. 
The antenna gain reduces drastically when an antenna configured for the far-field is used for communicating in the Fresnel region where there are phase variations \cite{Kay1960a}, as recently demonstrated experimentally in \cite{Tang2021a}, where this mismatched operation is called broadcasting.
The loss in gain can be recovered by assigning an elliptical phase-response over the aperture (e.g., implemented using an antenna array), which is tuned to the particular propagation distance of the receiver. 
The resulting beamforming has a beam width (BW) that is not only limited in the transverse angular domain, as in the far-field, but also in the depth domain \cite{Nepa2017a}. 
This feature is available in both the near-field and Fresnel region, thus IEEE nowadays recommends calling both the near-field \cite{Hansen1985a,IEEEstd145-2013}.
However, when operating adaptive antenna arrays, it remains important to distinguish between the physical effects that dominate at different propagation distances \cite{Friedlander2019a}.
This is particularly important since the communication and signal processing communities are now considering building base stations with large active antenna arrays for beyond 5G systems \cite{Bjornson2019d,Torres2020a} and design \emph{reconfigurable intelligent surfaces (RIS)} that can actively tune how signals are reflected \cite{BS20,Huang2018a,Wu2018a}.
In the latter case, the dominant use case might be to reflect signals to user devices in the Fresnel region and even in the near-field region (according to the traditional definition). The corresponding near-field beamforming gains were first characterized in \cite{Bjornson2019d} with some recent extensions in \cite{Garcia2020a,Feng2021a,Lu2021a}.

The purpose of this paper is to provide a primer on near-field channel modeling for active antenna arrays and RIS, using communication terminology.
Section~\ref{sec:passive-antenna} characterizes the essential regions for a single passive antenna. Section~\ref{sec:antenna-array} considers an antenna array and finite-depth near-field beamforming, while Section~\ref{sec:RIS} extends the analysis to an RIS. Different from the classical works \cite{Polk1956a,Kay1960a,Sherman1962a,Hansen1985a,Nepa2017a}, we use models from \cite{BS20,Dardari2020a} that take polarization into account.
The code is available at: \url{https://github.com/emilbjornson/nearfield-primer}

\section{Near-Field Region of a Passive Antenna}
\label{sec:passive-antenna}

A transmit antenna contains multiple point sources that emit polarized spherical waves.
The electric field at a distance $z$ from a point source, in any direction perpendicular to the propagation direction, is proportional to \cite[Eqs.~(3)-(4)]{Dardari2020a}
\begin{equation} \label{eq:Green-function-polarization}
 \frac{ \imagunit \eta e^{-\imagunit \frac{2\pi}{\lambda} z }  }{2 \lambda z} \left( 1 + \frac{\imagunit}{2\pi z / \lambda} - \frac{1}{(2\pi z / \lambda)^2}  \right)
\end{equation} 
where $\eta$ is the impedance of free-space and $\lambda$ is the wavelength. The last two terms are often neglected  since they decay  rapidly with the distance $z$ and thus are only influential in the \emph{reactive near-field}, very close to the antenna.
More precisely, the squared magnitude of the parenthesis in \eqref{eq:Green-function-polarization} is
\begin{equation}
\left| 1 + \frac{\imagunit}{2\pi z / \lambda} - \frac{1}{(2\pi z / \lambda)^2} \right|^2 = 
1 -  \frac{1}{(2\pi z / \lambda)^2}  +  \frac{1}{(2\pi z / \lambda)^4}
\end{equation}
which equals $0.975$ at distance $z= \lambda$. Hence, when considering an electrically small antenna that can be approximated as a single point source, the last two terms in \eqref{eq:Green-function-polarization} can be neglected when studying the electric field at distances $z \geq \lambda$.
For electrically large antennas consisting of a continuum of point sources with maximum length $D$, the reactive near-field ends approximately at $z = 0.62 \sqrt{D^3/\lambda}$ \cite{Selvan2017a}.

The analysis in the remainder of this paper will only consider propagation distances beyond the reactive near-field.

\subsection{Radiative near-field and Fresnel region}

Although a transmit antenna emits a spherical wave, the wave may locally appear as planar when observed at a sufficient distance. 
This happens in the so-called \emph{far-field} where the electric fields induced by the multiple point sources in the transmitter aperture superimpose to create
a field strength that is inversely proportional to the propagation distance $z$ (as in the first term in \eqref{eq:Green-function-polarization}) with a proportionality constant only depending on the angle \cite{Friedlander2019a}.
There is no notion of receive antenna in the standard far-field terminology \cite{IEEEstd145-2013}, but follows immediately that an ideal isotropic receive antenna will exhibit a wireless channel with the corresponding pathloss properties.

\begin{figure}[t!]
	\centering
	\begin{overpic}[width=\columnwidth,tics=10]{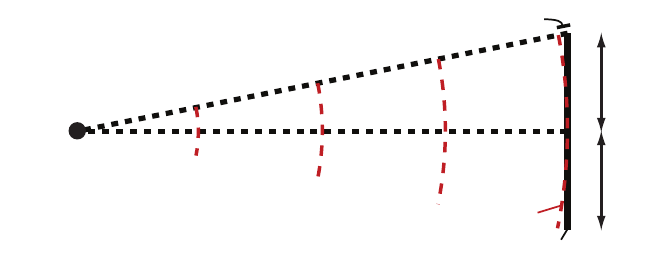}
	 \put (0,15) {Transmitter}
	 \put (75,0) {Receiver}
	 \put (63,6) {Wavefront}
	 \put (51,30) {$d'$}
	 \put (51,15) {$d$}
	 \put (71.5,36) {$d'\!-\!d$}
	 \put (94,11.5) {$\frac{D}{2}$}
	 \put (94,26.5) {$\frac{D}{2}$}
\end{overpic} 
	\caption{The curvature of an impinging spherical wave creates a delay $d'\!-\!d$ between the center of the receiver and the edge. The delay turns into a phase-shift of $\frac{2\pi}{\lambda} (d'\!-\!d)$.}
	\label{fig:fraunhofer}  \vspace{-3mm}
\end{figure}

To characterize the far-field, it is instructive to consider the opposite setup \cite{Selvan2017a}: an isotropic transmit antenna and an arbitrary receive antenna with some  maximum length $D$. Due to reciprocity, in the far-field, the amplitude of the electric field is constant over the receive antenna and the phase variation only depends on the incident angle, not the distance.
It is commonly assumed that the far-field
begins at the distance
\begin{equation} \label{eq:Fraunhofer}
d_F = \frac{2D^2}{\lambda}
\end{equation}
 from the transmitter \cite{Selvan2017a}. The distance $d_F$ is called the \emph{Fraunhofer distance} \cite{Sherman1962a} and is often motivated by considering the phase difference between the center and the corner caused by the wave's curvature. The largest variation occurs when the wave impinges perpendicularly, as shown in Fig.~\ref{fig:fraunhofer}.
 If the center is at the Fraunhofer distance $d= d_F$, then $d'$ in Fig.~\ref{fig:fraunhofer} is $d' = \sqrt{d_F^2 + (D/2)^2}$. Hence, the phase difference is
\begin{equation} \label{eq:Fraunhofer-phase}
\frac{2\pi}{\lambda} \left( \sqrt{d_F^2 + \frac{D^2}{4}} - d_F\right) \approx 
\frac{2\pi}{\lambda}  \frac{D^2}{8 d_F}  = \frac{\pi}{8},
\end{equation}
which is negligible when analyzing antenna patterns. 
We used the Taylor approximation $\sqrt{1+x} \approx 1+\frac{x}{2}$  for $x=\frac{D^2}{4d_F^2}$ in  \eqref{eq:Fraunhofer-phase}.
The approximation error is smaller than $3.5 \cdot 10^{-3}$  if $d_F \geq 1.2D$, which sets an additional lower limit on the far-field region \cite{Sherman1962a}.  This limit represents an angular difference between the center and edge of at most $\pi/8$, leading to a negligible amplitude difference  $d/d' \geq 1.2D/\sqrt{(1.2D)^2+D^2/4} \approx \cos(\pi/8) \approx 0.92$. 
The range of propagation distances between $1.2D$ and $d_F$ is called the \emph{Fresnel region} and is characterized by the fact that the amplitude variations can be neglected, but not the phase variations. The Fresnel region only exists if $d_F \geq 1.2D$, which implies $D \geq 0.6 \lambda$, thus the Fraunhofer distance is only applicable to electrically large antennas.

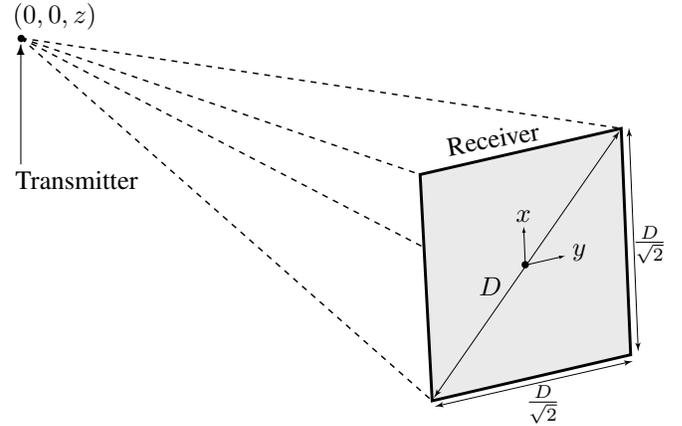
\begin{figure}[t!]
\begin{center}
	\begin{overpic}[width=\columnwidth,tics=10]{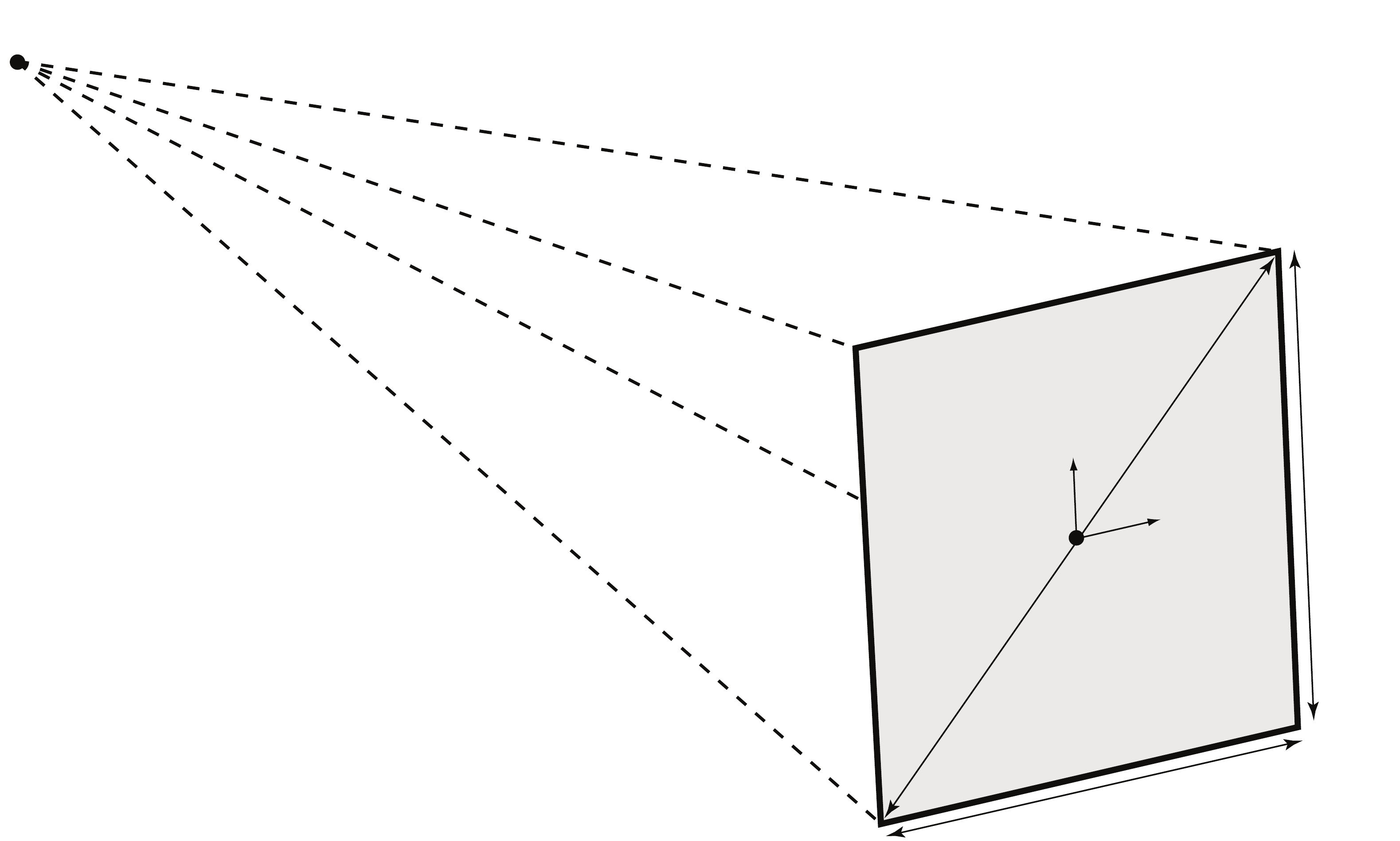}
\put(65,38){\rotatebox{10}{Receiver}}
\put(1.2,36.7){\vector(0,1){18}}
		\put(0.4,33){Transmitter}
		\put(93,24){$\frac{D}{\sqrt{2}}$}
		\put(77,0){$\frac{D}{\sqrt{2}}$}
		\put(70,17){$D$}
		\put(0,58){$(0,0,z)$}
		\put(75.5,28){$x$}
		\put(84,23){$y$}
\end{overpic} 
\end{center} 
\caption{A point source transmits to a receive antenna centered at the origin in the $xy$-plane and having with diagonal $D$.} \label{figure_geometric_fraunhofer}  
\end{figure}

We will provide a free-space propagation example to analyze the behavior at shorter path distances than $d_F$.
We consider an isotropic transmitter  at $(0,0,z)$ and a planar receive antenna covering $\mathcal{A} = \{ (x,y,0) : |x| \leq D/\sqrt{8}, |y| \leq D/\sqrt{8} \}$ in the $xy$-plane, as illustrated in Fig.~\ref{figure_geometric_fraunhofer}. 
When a signal with wavelength $\lambda$ and electric intensity $E_0$ (measured in Volt) is emitted with polarization in the $y$-dimension, the electric field perpendicular to a receive antenna at location $(x,y,0)$ is obtained as \cite[App.~A]{BS20}, \cite{Dardari2020a}
\begin{equation} \label{eq:intensity-function}
E(x,y) = \frac{E_0}{\sqrt{4 \pi}} \frac{\sqrt{z (x^2 + z^2)}}{(x^2+y^2+z^2)^{5/4}} e^{-\imagunit \frac{2\pi}{\lambda}\sqrt{x^2+y^2+z^2}},
\end{equation}
where $\sqrt{x^2+y^2+z^2}$ is the Euclidean distance.
The received power can then be computed as $E_0^2 |h|^2 / \eta$, where the dimensionless complex-valued channel response to the receive antenna is \cite[Eq.~(64)]{BS20}
\begin{equation}
h = \frac{1}{E_0} \sqrt{\frac{2}{D^2}} \int_{A} E(x,y) dx dy
\end{equation}
and $\frac{D^2}{2}$ is the area of the receive antenna.
The receive antenna gain is traditionally defined as \cite[Eq.~(6)]{Kay1960a}
\begin{equation}
G =  \frac{ \left| \int_{\mathcal{A}}  E(x,y) dx \, dy \right|^2}{\frac{\lambda^2}{4\pi} \int_{\mathcal{A}} \left| E(x,y) \right|^2 dx \, dy  }, \label{eq:G-exact}
\end{equation}
where $\frac{\lambda^2}{4\pi}$ is the area of an isotropic antenna. 
The gain in \eqref{eq:G-exact} becomes $G_{\textrm{plane}} = \frac{4\pi}{\lambda^2} \frac{D^2}{2}$ for an incident perpendicular plane wave with $E_{\textrm{plane}}(x,y) = E_0/(\sqrt{4\pi} z)$, which is the largest gain and achievable in the far-field.
We want to study the \emph{relative} antenna gain difference occurring in the near-field or Fresnel region (as compared to $G_{\textrm{plane}}$), thus we define the \emph{normalized antenna gain} as 
\begin{equation}
G_{\textrm{antenna}} =  \frac{G}{G_{\textrm{plane}}} =  \frac{ \left| \int_{\mathcal{A}}  E(x,y) dx \, dy \right|^2}{\frac{D^2}{2} \int_{\mathcal{A}} \left| E(x,y) \right|^2 dx \, dy  }. \label{eq:G-exact-normalized}
\end{equation}
This is the received power $E_0^2 |h|^2 / \eta$
normalized by the total power $\int_{\mathcal{A}} \left| E(x,y) \right|^2 dx \, dy / \eta$ over the antenna aperture.

The exact gain~\eqref{eq:G-exact-normalized} can be computed by numerical integration of the electric field in \eqref{eq:intensity-function}. An analytical expression can be obtained using that $E(x,y) \approx \frac{E_0}{\sqrt{4\pi} z} e^{-\imagunit \frac{2\pi}{\lambda} (z + \frac{x^2}{2z} + \frac{y^2}{2z})}$ for $z \geq 1.2D$.
This classic Fresnel approximation leads to \cite[Eq.~(22)]{Polk1956a}
\begin{align} \notag
G_{\textrm{antenna}} &\approx \left( \frac{2}{D^2} \right)^2 \left| e^{-\imagunit \frac{2\pi}{\lambda} z  }\int_{\mathcal{A}}  e^{-\imagunit \frac{2\pi}{\lambda} (\frac{x^2}{2z} + \frac{y^2}{2z})} dx \, dy  \right|^2 \\ \notag
&= \left( \frac{2}{D^2} \right)^2 \left| \int_{-D/\sqrt{8}}^{D/\sqrt{8}}  e^{-\imagunit \frac{\pi}{\lambda} \frac{x^2}{z}} dx \right|^4 \\
&= \left( \frac{8 z}{d_F} \right)^2 \left( C^2 \left( \sqrt{\frac{d_F}{8 z} }\right) \!+\! S^2 \left( \sqrt{\frac{d_F}{8 z} }\right) \right)^2 \label{eq:G-approx},
\end{align}
where $C(\cdot)$ and $S(\cdot)$ are the Fresnel integrals.\footnote{We use the Fresnel integral definitions $C(x) = \int_{0}^{x} \cos(\pi t^2/2) dt$ and $S(x) = \int_{0}^{x} \sin(\pi t^2/2) dt$. These can be computed using the error function.}

\begin{figure}[t!]
\begin{center}
	\begin{overpic}[trim={5mm 5mm 12mm 5mm},width=\columnwidth,tics=10]{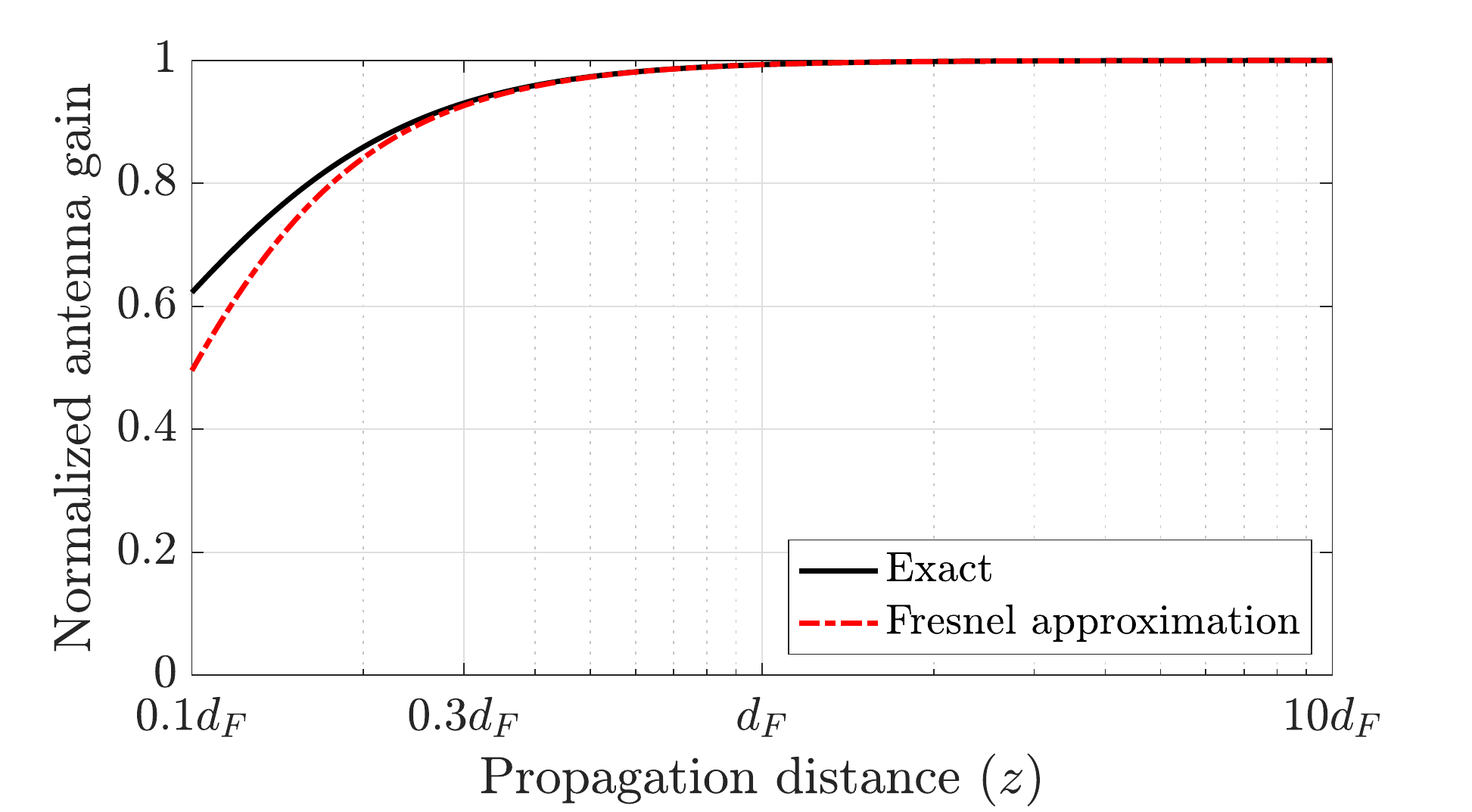}
\end{overpic} 
\end{center} 
\caption{The normalized antenna gain reduces when the propagation distance is shorter than the Fraunhofer distance $d_F$.} \label{figure_fraunhofer_antenna}  
\end{figure}

Fig.~\ref{figure_fraunhofer_antenna} shows the normalized antenna gain $G_{\textrm{antenna}}$ as a function of the distance $z$ from the transmitter to the center of the receiver. The antenna diagonal is $D=2\lambda$, which gives $d_F = 8 \lambda = 4 D$, thus the Fresnel region is $0.3 d_F \leq z \leq d_F$ according to definition from \cite{Sherman1962a}. 
 We consider the exact gain computed numerically using \eqref{eq:G-exact-normalized} and the approximate expression in \eqref{eq:G-approx}.
For $z \geq d_F$, the curvature of the impinging wave is so weak that at least $99\%$ of the maximum gain $G_{\textrm{plane}}$ is achieved.
The gain drops as $z$ is reduced, reaching $93\%$ at $z=0.3d_F = 1.2 D$. The approximation in  \eqref{eq:G-approx} is accurate for $z \geq 0.3 d_F$, 
confirming it as the Fresnel region's lower limit. Since the Fresnel approximation neglects polarization effects, it has no dominant impact when considering a single antenna.

In summary, the Fraunhofer distance is an accurate lower limit of the far-field of an antenna. 
From a strict electromagnetic perspective, the far-field can only be measured from a transmitter without considering the existence of a receiver. However, for communication channels, we can measure it in either direction, as long as the transmitter or receiver is isotropic. An isotropic transmitter is said to be in the far-field of a receiver, if the phase and amplitude variations over the aperture can be neglected when computing the antenna gain.

\section{Near-Field Region of an Antenna Array}
\label{sec:antenna-array}

We will now consider the near-field and Fresnel region of a planar square array of $N$ identical antennas, where $\sqrt{N}$ is an integer for simplicity. 
As in the previous section, we will take a receiver perspective and assume an isotropic transmitter, but the same result holds in the reciprocal setup with a transmitting array \cite{Sherman1962a,Polk1956a,Kay1960a,Hansen1985a}.
The antennas are of the same kind as in the previous section and deployed edge-to-edge on a square grid in the $xy$-plane. We let $n \in \{1,\ldots,\sqrt{N}\}$ denote the row-index in the $x$ dimension and $m \in \{1,\ldots,\sqrt{N}\}$ denote the column index in the $y$ dimension. The antenna with index $(n,m)$ is then centered at the point $(\bar{x}_n,\bar{y}_m,0)$ given by
\begin{equation}
\bar{x}_n = \Big(n-\frac{\sqrt{N}+1}{2} \Big) \frac{D}{\sqrt{2}}, \,\, \bar{y}_m = \Big(m-\frac{\sqrt{N}+1}{2} \Big) \frac{D}{\sqrt{2}}
\end{equation}
and covers the area
\begin{align} 
\!\!\mathcal{A}_{n,m} = \left\{ (x,y,0) : \left| x - \bar{x}_n \right| \leq \frac{D}{\sqrt{8}},  \left| y - \bar{y}_m  \right| \leq \frac{D}{\sqrt{8}} \right\}.
\end{align}
For an impinging wave with $E(x,y)$ in \eqref{eq:intensity-function}, the complex-valued channel response to receive antenna $(n,m)$ is
\begin{equation}
h_{n,m} = \frac{1}{E_0} \sqrt{\frac{2}{D^2}} \int_{\mathcal{A}_{n,m}}  E(x,y) dx \, dy.
\end{equation}
If an information symbol $s \in \mathbb{C}$ with power $E_0^2/\eta$ is modulated onto the waveform, the received signal at antenna $(n,m)$ is
\begin{equation}
r_{n,m} = h_{n,m} s + w_{n,m}, \quad n,m \in \{1,\ldots,\sqrt{N}\},
\end{equation}
where $w_{n,m} \sim \CN(0,\sigma^2)$ is independent complex Gaussian receiver noise.
This is a single-input multiple-output (SIMO) channel, thus the signal-to-noise ratio (SNR) is maximized using matched filtering where $c_{n,m} = h_{n,m}^*$ is the weight assigned to antenna $(n,m)$. The maximum SNR is
\begin{equation} \label{eq:SNR-formula}
\mathrm{SNR} = 
 \frac{E_0^2}{\eta \sigma^2} \sum_{n=1}^{\sqrt{N}} \sum_{m=1}^{\sqrt{N}}  \left| h_{n,m} \right|^2.
\end{equation}

Inspired by \eqref{eq:G-exact-normalized}, we define the normalized \emph{antenna array gain} as 
\begin{equation} \label{eq:antenna-array-gain-exact}
G_{\textrm{array}} = \frac{ \sum_{n=1}^{\sqrt{N}} \sum_{m=1}^{\sqrt{N}}  \left|  \int_{\mathcal{A}_{n,m}}  E(x,y) dx \, dy \right|^2 }{ N \frac{D^2}{2}  \int_{\mathcal{A}} \left| E(x,y) \right|^2 dx \, dy},
\end{equation}
which is the total received power $E_0^2 \sum_{n,m} | h_{n,m}|^2/ \eta$ from \eqref{eq:SNR-formula} 
normalized by the total power $N \int_{\mathcal{A}} \left| E(x,y) \right|^2 dx \, dy / \eta$ impinging on $N$ reference antennas (as if they were all in the origin).
This new definition captures both the antenna and array gains since these only decouple in the far-field.

We can either compute \eqref{eq:antenna-array-gain-exact} numerically or use the Cauchy-Schwarz inequality to obtain a closed-form upper bound as:\footnote{We use
$ |  \int_{\mathcal{A}_{n,m}}  E(x,y) dx  dy |^2 \!\leq \!   \int_{\mathcal{A}_{n,m}}  |E(x,y)  |^2 dx  dy \int_{\mathcal{A}_{n,m}}1 dx  dy$.}
\begin{align} \notag
& G_{\textrm{array}}  \leq  \frac{ \sum_{n=1}^{\sqrt{N}} \sum_{m=1}^{\sqrt{N}}  \frac{D^2}{2} \int_{\mathcal{A}_{n,m}}  \left| E(x,y) \right|^2 dx \, dy  }{ N  \frac{D^2}{2}     \int_{\mathcal{A}} \left| E(x,y) \right|^2 dx \, dy} \\ \notag
& =  \frac{  \int_{-D\sqrt{N/8}}^{D\sqrt{N/8}}  \int_{-D\sqrt{N/8}}^{D\sqrt{N/8}}  \left| E(x,y) \right|^2 dx \, dy  }{  N  \int_{-D/\sqrt{8}}^{D/\sqrt{8}}  \int_{-D/\sqrt{8}}^{D/\sqrt{8}} \left| E(x,y) \right|^2 dx \, dy} \\
& \overset{(a)}{=} \frac{ 
 \frac{N \alpha_z  }{2(N \alpha_z +1) \sqrt{2 N \alpha_z  + 1}} + \tan^{-1} \!\left( \frac{N \alpha_z }{ \sqrt{2N \alpha_z  + 1}} \right) 
}{ \frac{ N \alpha_z  }{2(\alpha_z +1) \sqrt{2 \alpha_z  + 1}} + N  \tan^{-1} \!\left( \frac{\alpha_z }{ \sqrt{2 \alpha_z  + 1}} \right) } \label{eq:G-upper-bound},
\end{align}
where $\alpha_z = \frac{D^2}{8 z^2}$ and $(a)$ follows from \cite[Cor.~1]{BS20}.
The upper bound in \eqref{eq:G-upper-bound} is tight when $E(x,y)$ is approximately constant over each individual antenna, which happens for electrically small antennas (i.e., $D \ll \lambda$) and in the far-field.

The antenna array gain depends on the propagation distance and the goal of this section is to determine when the maximum normalized gain is achievable.
A partial answer can be obtained by revisiting the Fraunhofer distance in \eqref{eq:Fraunhofer}. The maximum length of the considered array is the diagonal $D \sqrt{N}$. If we want the spherical curvature to cause negligible phase variations over the array's diagonal, we need $z \geq d_{FA}$, where we define the \emph{Fraunhofer array distance} as
\begin{equation}
d_{FA} = \frac{2 \left(D \sqrt{N} \right)^2}{\lambda} = N d_F.
\end{equation}
We can notice that it is precisely $N$ times larger than the Fraunhofer distance of an individual antenna.
When analyzing the far-field, we must not forget the additional condition that the amplitude variations must be negligible over the array.
For the considered case with polarized waves, \cite{BS20} noticed that the amplitude variations are negligible for $z \geq d_B$, where
\begin{equation}
d_B = 2 D \sqrt{N}
\end{equation}
is twice array's length. We call this the \emph{Bj\"ornson distance}
and notice that it only grows with the square root $\sqrt{N}$ of the number of antennas and, thus, will be much shorter than $d_{FA}$ for arrays with many antennas. 
To identify how large $N$ must be for this to happen, we rewrite the condition $d_{FA} \geq d_B$ as
\begin{equation}
\frac{2 \left(D \sqrt{N} \right)^2}{\lambda}  \geq 2 D \sqrt{N}  \,\,\, \Rightarrow \,\,\, N \geq \frac{\lambda^2}{D^2}.
\end{equation}
It holds for $N\geq 1$ if $D=\lambda$ and for $N \geq 16$ if $D=\lambda/4$. 
The Bj\"ornson distance is larger than $d_F$ whenever $N \geq (D/\lambda)^2$.

\begin{figure}[t!]
\begin{center}
	\begin{overpic}[trim={5mm 5mm 12mm 5mm},width=\columnwidth,tics=10]{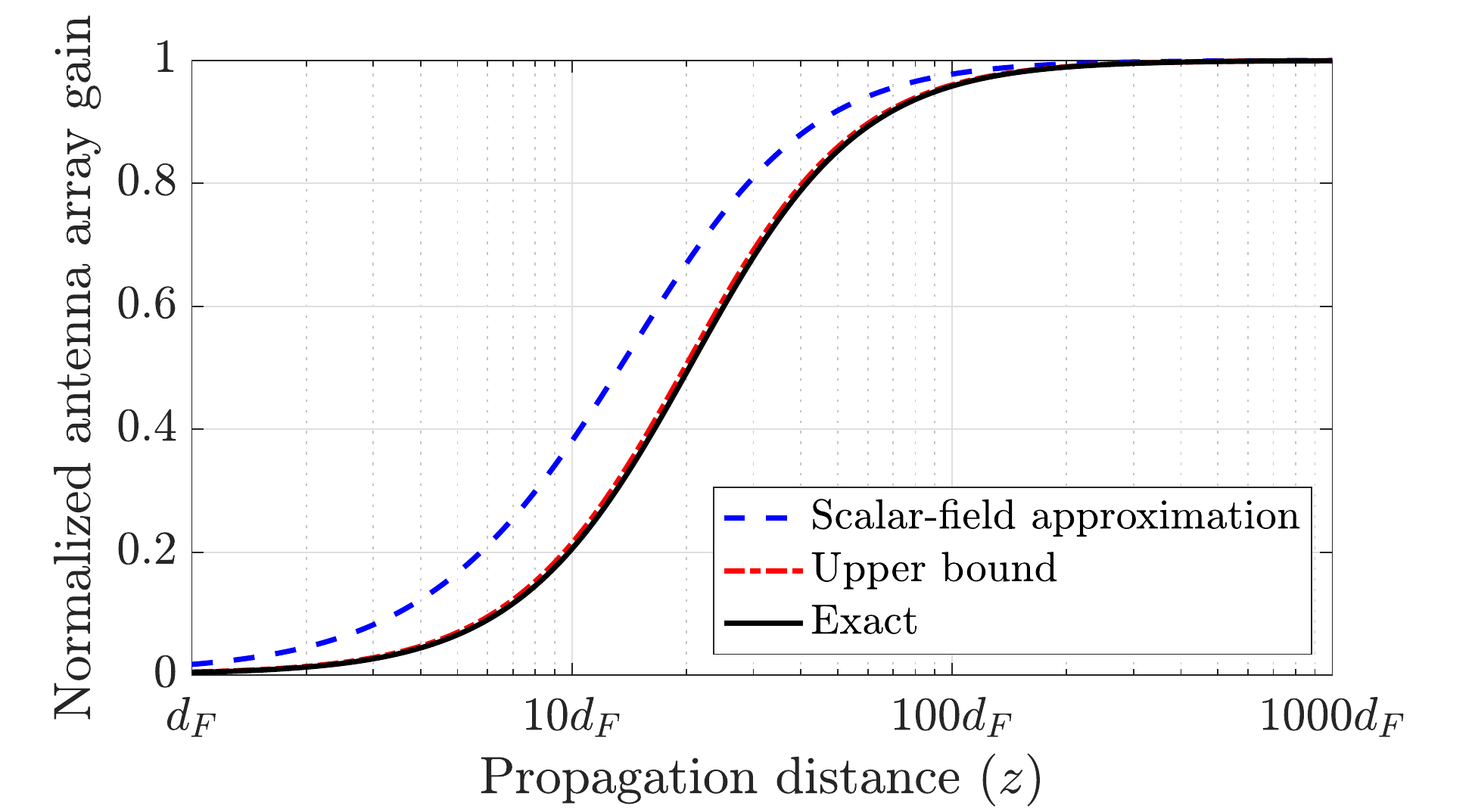}
	 \put(68,47){\vector(0,1){4}}
	 \put(66,44){\footnotesize $d_{B}$}
	 \put(90.4,49){\vector(0,1){4}}
	 \put(87,46){\footnotesize $d_{FA}$}
\end{overpic} 
\end{center} 
\caption{The normalized antenna array gain is below its maximum when the propagation distance is shorter than $d_B$.} \label{figure_fraunhofer_array}  \vspace{-2mm}
\end{figure}
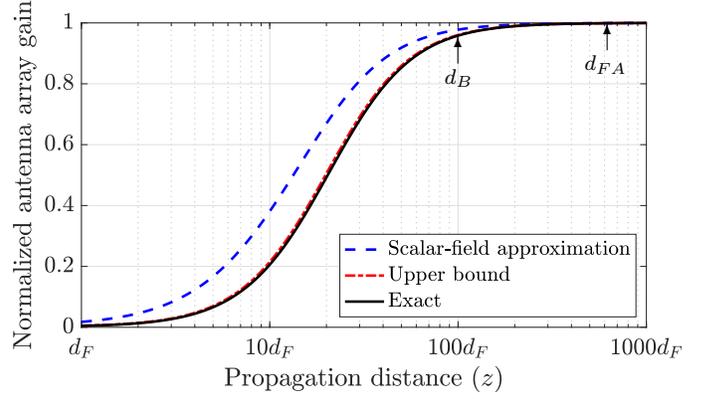

Fig.~\ref{figure_fraunhofer_array} shows the normalized antenna array gain $G_{\textrm{array}} $ as a function of the distance $z$ from the transmitter to the center of the receiver.
We consider an array of $N=25^2=625$ antennas, each with length $D=\lambda/4$.
The Fraunhofer distance becomes $d_F = \lambda/8$ and is used as a reference unit on the horizontal axis.
In this case, we get $d_{FA} = 625 d_F$ and $d_B = 100 d_F$.
We compare the exact $G_{\textrm{array}}$ computed using \eqref{eq:antenna-array-gain-exact} and the upper bound in \eqref{eq:G-upper-bound}.
At least 95\% of the maximum gain is achieved for $z \geq d_B$, while the Fraunhofer array distance $d_{FA}$ has no evident impact on the curves. The bound in \eqref{eq:G-upper-bound} is tight since each antenna is electrically small.

The electric field in \eqref{eq:intensity-function} differs from the classic scalar-field approximation \cite{Sherman1962a}
$\frac{E_0}{\sqrt{4\pi} \sqrt{x^2+y^2+z^2}} e^{-\imagunit \frac{2\pi}{\lambda}\sqrt{x^2+y^2+z^2}}$ that neglects how the polarization losses and effective antenna areas vary with the incident angle over the array.
Fig.~\ref{figure_fraunhofer_array} shows the curve that is obtained using this approximation and it is only tight for $z \geq d_B$. Hence, it is clear that the varying polarization losses and effective areas must be considered when analyzing $z < d_B$ (as previously noted in \cite{BS20}).

\subsection{Is the Fraunhofer array distance irrelevant?}

The example shows that we (almost) achieve the maximum antenna array gain whenever the receiver is at a distance $z \geq d_B \geq d_F$ larger than the Bj\"ornson distance, even if $z \leq d_{FA}$. Hence, the Fraunhofer array distance is not characterizing this operating regime.
The reason is that each individual antenna is in the far-field of the transmitter \cite{Friedlander2019a} and, thus, observes a locally plane wave that it can extract the maximum gain from.
If $d_B \leq z \leq d_{FA}$ (i.e., the array-equivalent of the Fresnel region), the spherical curvature is  noticeable when comparing local phases between antennas, even if the wave is locally plane at each antenna.
The matched filtering compensates for this phase difference.
This is a key difference between having a single large antenna and an equal-sized array of smaller antennas: the latter can compensate for the phase variations, while the former exhibits a gain loss since it can not.

Nevertheless, the Fraunhofer array distance is not irrelevant but characterizes what kind of receiver processing is needed and the resulting depth-of-focus (DF). If $z \geq d_{FA}$, we can use the plane wave approximation to determine the array response vector based only on the incident angle and use it for matched filtering. All signals arriving from that angle will be amplified.
In contrast, for $z \leq d_{FA}$, we need to consider the spherical curvature when computing the matched filtering weights $c_{n,m}$ and thereby know the transmitter distance more precisely.

The receiving array has a DF that shrinks the closer the transmitter is, just as the DF of a camera shrinks when focusing on a nearby object.
When the matched filtering is selected to focus on a potential transmitter at distance $z = F$, the DF can be defined as the distance interval $z \in [F_{\min},F_{\max}]$ where the antenna array gain is at most $3$\,dB lower than the maximum value \cite{Sherman1962a,Nepa2017a}. 

We can compute the DF for an array with electrically small antennas using the Fresnel approximation in \eqref{eq:G-approx} and inject the phase-shift $e^{+\imagunit \frac{2\pi}{\lambda} (\frac{x^2}{2F} + \frac{y^2}{2F})}$ representing (continuous) matched filtering into the integral, integrate over the entire area of the array, and replace $2/D^2$ by $2/(D^2N)$. We then 
obtain the approximate normalized antenna array gain tuned for $z=F$:
\begin{align} \notag
& \left( \frac{2}{D^2 N} \right)^{\!2} \! \left| e^{-\imagunit \frac{2\pi}{\lambda} z  } \! \int_{- D\sqrt{N/8}}^{D\sqrt{N/8}} \int_{- D\sqrt{N/8}}^{D\sqrt{N/8}}  \!  e^{-\imagunit \frac{2\pi}{\lambda} (\frac{x^2}{2z_{\textrm{eff}}} + \frac{y^2}{2z_{\textrm{eff}}})} dx \, dy  \right|^2 \\ 
&=
\left( \frac{8 z_{\textrm{eff}}}{d_{FA} } \right)^2 \left( C^2 \left( \sqrt{\frac{d_{FA} }{8 z_{\textrm{eff}}} }\right) + S^2 \left( \sqrt{\frac{d_{FA}}{8 z_{\textrm{eff}}} }\right) \right)^2 \label{eq:G-approx-DF},
\end{align}
where $z_{\textrm{eff}} = \frac{Fz}{|F-z|}$ represents the focal point deviation. 
Note that  \eqref{eq:G-approx-DF} has the structure  $A(x) = (C^2(\sqrt{x}) + S^2(\sqrt{x}) )^2 /x^2$, where $x = d_{FA}  / (8 z_{\textrm{eff}})$. Moreover, $A(x) $ is a decreasing function for $x \in [0,2]$ with $A(0) =1$ and $A(1.25) \approx 0.5$.  Hence, the $3$\,dB gain loss is obtained when 
\begin{equation}
1.25 = \frac{Nd_F}{8 z_{\textrm{eff}}} = \frac{Nd_F |F-z|}{8 Fz}  \,\, \rightarrow \,\, z = \frac{d_{FA} F  }{d_{FA}   \pm 10 F}.
\end{equation}

\begin{theorem} \label{th:DF}
When matched filtering is utilized to focus on a transmitter at distance $F$, the DF is
\begin{equation}
z \in \left[ \frac{d_{FA} F }{d_{FA}  + 10 F} ,  \frac{d_{FA} F  }{d_{FA}  - 10 F} \right] \label{eq:BD-interval}
\end{equation}
if $F < d_{FA} /10$. Otherwise, the upper limit is replaced by $\infty$.
\end{theorem}

Based on this novel result, we conclude that beamforming with a limited DF is only achieved when focusing on a point closer than $d_{FA} /10$. For more distant points, the $3$\,dB beam depth (BD) extends to infinity, as in conventional far-field beamforming. Moreover, as the focal point $F \to \infty$, the lower limit approaches $d_{FA} /10$, making it the clear border between near-field and far-field beamforming. We define the length of the interval in Theorem~\ref{th:DF} as the \emph{3\,dB BD} and it is given as
\begin{align}
    \textrm{BD}_{\textrm{3dB}} = \begin{cases} \frac{d_{FA} F  }{d_{FA}  - 10 F}-\frac{d_{FA} F  }{d_{FA}  + 10 F} = \frac{20 d_{FA} F^2}{d_{FA}^2 - 100F^2}, & F < \frac{d_{FA}}{10}, \\
    \infty,  & F \geq \frac{d_{FA}}{10}.
    \end{cases}
\end{align}

\begin{figure}[t!]
\begin{center}
	\begin{overpic}[trim={5mm 5mm 12mm 5mm},width=\columnwidth,tics=10]{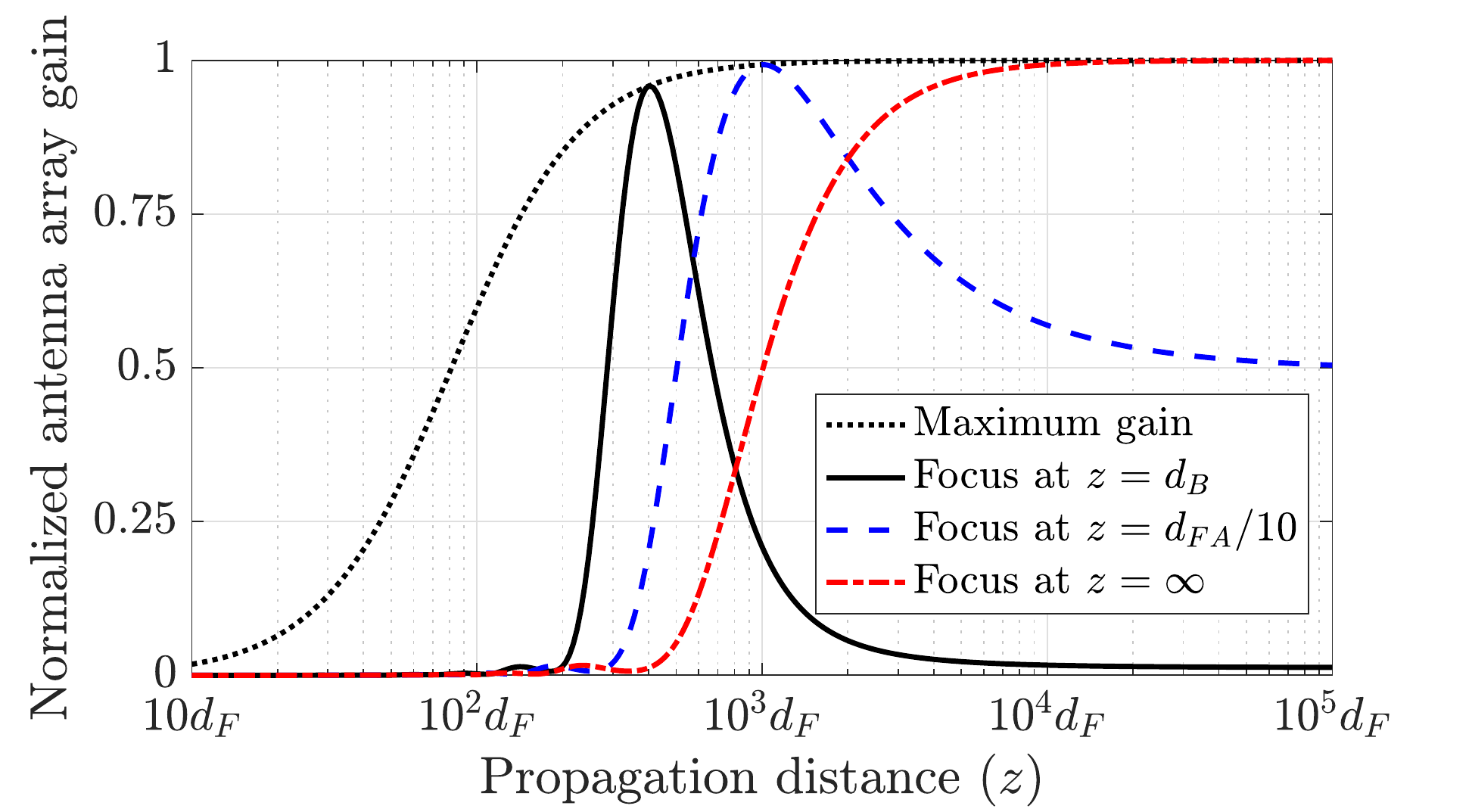}
	 \put(75.1,48){\vector(0,1){4}}
	 \put(72.1,45){\footnotesize $d_{FA}$}
	 \put(53.9,48){\vector(0,1){4}}
	 \put(50.7,45){\footnotesize $\frac{d_{FA}}{10}$}
	 \put(43.5,54){\footnotesize $d_{B}$}
\end{overpic} 
\end{center} 
\caption{The depth of a beam depends on the location the matched filtering focuses on. The depth is finite for near-field beamforming, when the focal point is closer than $d_{FA}/10$.} \label{figure_focusing}   \vspace{-2mm}
\end{figure}
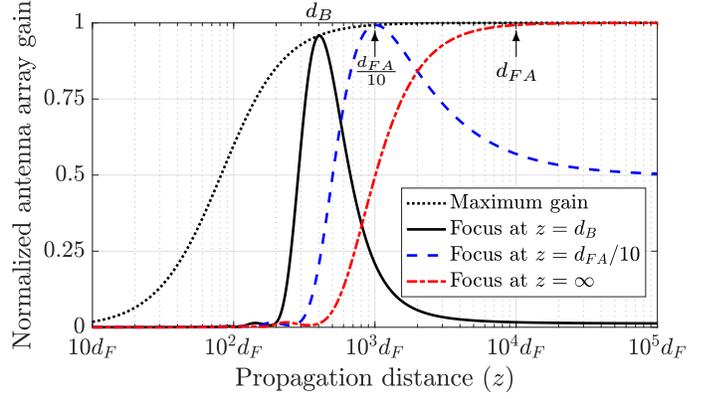

Fig.~\ref{figure_focusing} shows the maximum normalized gain that can be achieved at different distances  from an array with $N=100^2 = 10^4$ antennas, each with length $D = \lambda/4$.
The figure also shows how signals arriving from different distances are amplified when the matched filtering is selected to focus on a transmitter located at three different distances.
The Fraunhofer distance is $d_F = \lambda/8$, while $d_B = 400 d_F$ and $d_{FA} = 10^4 d_F$.
For far-field focusing on $z=\infty$, the normalized gain is between $1$ and $0.5$ (i.e., $-3$\,dB) in the interval $[d_{FA}/10,\infty)$. This is the DF and the observed interval is in line with Theorem~\ref{th:DF}.
If the matched filtering focus on $z = d_{FA}/10$, the DF is $[ d_{FA}/20,\infty)=[ 500 d_F ,\infty)$ but we approach $-3$\,dB as $z \to \infty$, which is why $d_{FA}/10$ is the largest distance for which the DF is infinite.
If we the matched filtering focus on $z = d_B$, the DF interval is roughly $[286 d_F, 667 d_F]$, which is rather narrow. As expected, there is also a noticeable loss in maximum gain when the focal point is the Bj\"ornson distance.

\section{Near-Field Region of an RIS}
\label{sec:RIS}

\begin{figure}[t!]
	\centering
	\begin{overpic}[width=\columnwidth,tics=10]{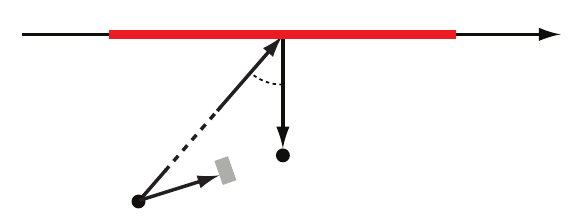}
	 \put (3,5.5) {Transmitter}
	 \put (51,14.5) {Receiver}
	 \put (0,-1) {$\left(-\rho \sin\left(\theta_i\right),0,\rho\cos\left(\theta_i\right)\right)$}
	 \put (43.5,20.5) {$\theta_i$}
	 \put (51,10) {$(0,0,z)$}
	 \put (94,27) {$x$}
	 \put (45,33.5) {RIS}
	 \put (7.5,34) {$-D\sqrt{N/8}$}
	 \put (70,34) {$D\sqrt{N/8}$}
\end{overpic}  \vspace{1mm}
	\caption{A transmitter communicates with a receiver via an RIS, while the direct path is blocked.}
	\label{fig:RIS_setup}  \vspace{-3mm}
\end{figure}
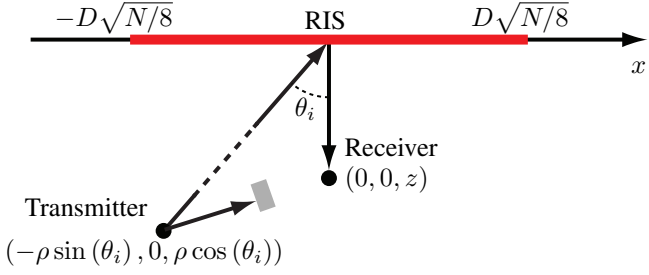

We now consider an RIS with $N$ reconfigurable elements with the same geometric array structure as in the previous section. It is deployed to beamform the impinging signal from a single-antenna transmitter to a single-antenna receiver, as shown in Fig.~\ref{fig:RIS_setup}.
The isotropic transmitter is located at $\left(-\rho \sin\left(\theta_i\right),0,\rho\cos\left(\theta_i\right)\right)$, where $\theta_i$ is the incident angle in the $xz$-plane and $\rho$ is the distance to the center of the RIS. The distance $\rho$ is assumed large enough so that the transmitter is in the far-field of the RIS.\footnote{For brevity, we are not covering the (somewhat contrived) case where both the transmitter and receiver are in the near-field region of the RIS.}
 The transmitter emits a signal with electric intensity $E_i$
 that is polarized in the $y$-dimension 
 and can be approximated as a plane wave due to the far-field assumption.
 The electric field distribution on the RIS surface is
\begin{equation} \label{eq:incident-electric-field} 
    E_t(x,y) = \frac{E_i}{\sqrt{4\pi} \rho} e^{-\imagunit\frac{2\pi}{\lambda} \left( \rho + \sin\left(\theta_i\right)x \right)}.
\end{equation}
Due to the space limitation, we assume there is an unobstructed line-of-sight from the transmitter to the RIS and from the RIS to the receiver, but no other paths between the transmitter and receiver.
The receive antenna is located at $(0,0,z)$, thus the function $E(x,y)$ in \eqref{eq:intensity-function} can be used to find the field at the receiver antenna induced by the reflected waves from the RIS. More precisely, we define $\tilde{E}(x,y) = E(x,y)/E_0$, where the dependence on $E_0$ is removed since we call the intensity $E_i$ in this section.

The complex-valued channel response from the transmitter to the receiver  through RIS element $(n,m)$ is
\begin{equation}
g_{n,m} = e^{-\imagunit\varphi_{n,m}} \sqrt{\frac{2}{D^2}} \int_{\mathcal{A}_{n,m}} \frac{E_t(x,y)}{E_i} \tilde{E}(x,y) dx \, dy,
\end{equation}
where $\varphi_{n,m}\in[0,2\pi)$ is the phase-shift introduced by the  element, which also represents the phase of the reflection coefficient in the lumped circuit equivalent of the element \cite{Abeywickrama2020a}. 

If an information symbol $s \in \mathbb{C}$ with power $E_i^2/\eta$  is modulated onto the incident waveform, the received signal is
\begin{equation} \label{eq:RIS-signal-model}
r =  \sum_{n=1}^{\sqrt{N}}\sum_{m=1}^{\sqrt{N}}g_{n,m} s + w,
\end{equation}
where $w\sim \CN(0,\sigma^2)$ is independent complex Gaussian receiver noise. The SNR is maximized 
when the phase-shifts  are selected as \cite{Wu2018a}
\begin{align}
    \varphi_{n,m} =  \angle{\int_{\mathcal{A}_{n,m}}  \tilde{E}(x,y) e^{-\imagunit\frac{2\pi}{\lambda} \left( \rho + \sin\left(\theta_i\right)x \right)} dx \, dy}. \label{eq:focusing-phase-shifts}
\end{align}
 The resulting maximum SNR is
\begin{equation} \label{eq:SNR-formula-RIS}
\mathrm{SNR} = 
 \frac{E_i^2}{\eta \sigma^2} \left(\sum_{n=1}^{\sqrt{N}} \sum_{m=1}^{\sqrt{N}}  \left| g_{n,m} \right| \right)^2.
\end{equation}
Inspired by \eqref{eq:G-exact-normalized}, we define the normalized \emph{RIS gain} as 
\begin{equation} \label{eq:antenna-RIS-gain-exact}
G_{\textrm{RIS}} = \frac{ \left(\sum\limits_{n=1}^{\sqrt{N}} \sum\limits_{m=1}^{\sqrt{N}}  \left|  \int_{\mathcal{A}_{n,m}}  E_t(x,y) \tilde{E}(x,y) e^{-\imagunit\varphi_{n,m}} dx \, dy \right|\right)^2 }{ N^2 \frac{D^2}{2}  \int_{\mathcal{A}} | E_t(x,y) \tilde{E}(x,y) |^2 dx \, dy}
\end{equation}
which is the received power $E_i^2 (\sum_{n=1}^{\sqrt{N}} \sum_{m=1}^{\sqrt{N}}  \left| g_{n,m} \right| )^2/ \eta$ in \eqref{eq:SNR-formula-RIS} normalized by the power  $N^2 \int_{\mathcal{A}} | E_t(x,y)\tilde{E}(x,y) |^2 dx \, dy / \eta$ that could ideally reach the receiver via $N$ RIS elements of the reference type deployed in the origin. The normalization makes $G_{\textrm{RIS}}  \in [0,1]$.

We will later illustrate that the RIS creates a beam focused on  $(0,0,z)$ with a predictable depth and width. To analyze this, we first consider the normalized RIS gain that can be measured by a receiver at location $(x_r,y_r,z_r)$, when the RIS is configured to focus on another point $(0,0,z)$. 
It is given by 
\begin{align} \label{eq:antenna-RIS-gain-arbitrary-receiver}
G_{\textrm{RIS},r} = &\frac{ \left|\sum\limits_{n=1}^{\sqrt{N}} \sum\limits_{m=1}^{\sqrt{N}}  \int_{\mathcal{A}_{n,m}} \!\! E_r(x,y) e^{-\imagunit\varphi_{n,m}}  dx \, dy \right|^2 \!}{ N^2 \frac{D^2}{2}  \int_{\mathcal{A}} \left| E_r(x,y) \right|^2 dx \, dy},
\end{align} 
where the phase-shifts $\varphi_{n,m}$ are selected as in \eqref{eq:focusing-phase-shifts} and we have modified the end-to-end electric field expression by replacing $ \tilde{E}(x,y)$ with a formula from \cite[App.~A]{BS20}  to get \vspace{-3mm}
\begin{align} \label{eq:intensity-function2}
E_r(x,y) = & \frac{E_i}{4\pi \rho} \frac{\sqrt{z_r \left(\left(x-x_r\right)^2 + z_r^2\right)}}{(\left(x-x_r\right)^2+\left(y-y_r\right)^2+z_r^2)^{5/4}} \nonumber\\
&\times e^{-\imagunit \frac{2\pi}{\lambda}\sqrt{\left(x-x_r\right)^2+\left(y-y_r\right)^2+z_r^2}} e^{-\imagunit\frac{2\pi}{\lambda} \left( \rho + \sin\left(\theta_i\right)x \right)}.
\end{align}
We can compute the DF obtained by an RIS with electrically small elements that focuses a beam on $(0,0,F)$. To this end, we return to the Fresnel approximation in \eqref{eq:G-approx}, inject the phase-shift $e^{+\imagunit \frac{2\pi}{\lambda} \left(\frac{x^2}{2F} + \frac{y^2}{2F}+\rho+\sin\left(\theta_i\right)x\right)}$ into the integrand, and integrate over the entire surface (as in \eqref{eq:G-approx-DF}) to  obtain a continuous approximation of the RIS' phase-shifts. The additional phase-shift $ \frac{2\pi}{\lambda}(\rho+\sin\left(\theta_i\right)x)$  is used by the RIS to fully compensate for the phase-shift variations of the incident plane wave, giving an RIS gain expression equal to \eqref{eq:G-approx-DF}. Hence, we get the same DF as in the matched filtering case for a SIMO channel. In other words, the re-radiated signal from the RIS is the same as if this was the original transmitter, except for a scaling factor representing the end-to-end propagation loss between the transmitter and receiver.

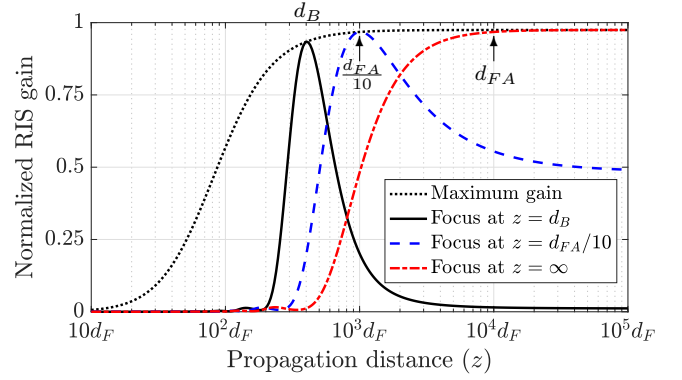
\begin{figure}[t!]
\begin{center}
	\begin{overpic}[trim={5mm 5mm 12mm 5mm},width=0.95\columnwidth,tics=10]{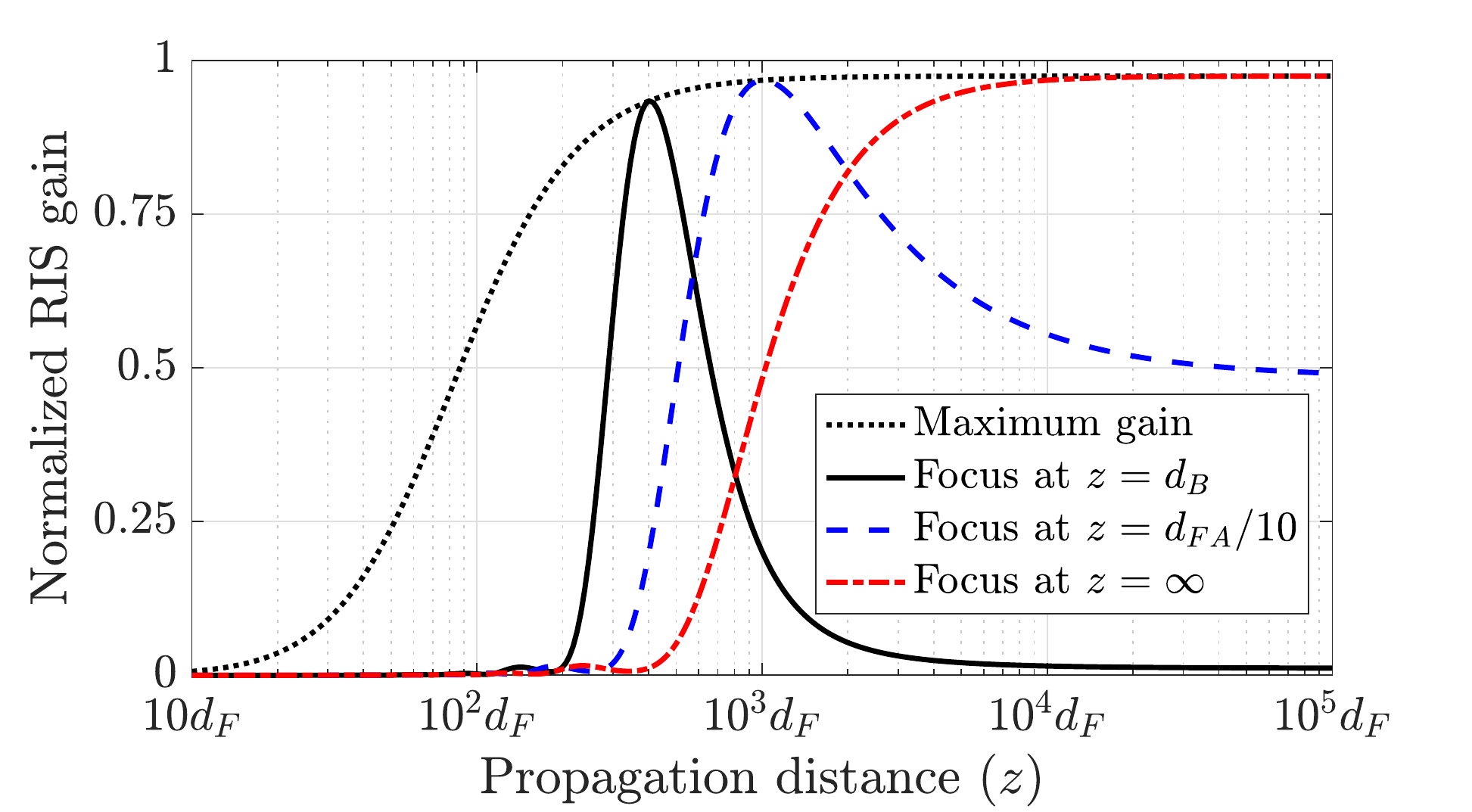}
	 \put(75.1,47){\vector(0,1){4}}
	 \put(72.1,44){\footnotesize $d_{FA}$}
	 \put(43.5,54){\footnotesize $d_{B}$}
	 \put(53.9,47){\vector(0,1){4}}
	 \put(50.7,44){\footnotesize $\frac{d_{FA}}{10}$}
\end{overpic} 
\end{center} 
\caption{The depth of a beam depends on the location it is focused on. The depth is finite for near-field beamforming, when the focal point is closer than $d_{FA}/10$.} \label{figure_focusing-RIS} \vspace{-2mm}
\end{figure}

In Fig.~\ref{figure_focusing-RIS}, we assume the  angle of the incident plane wave is $\theta_i=\pi/6$ and the RIS has  $N=100^2 = 10^4$ elements, each with length $D = \lambda/4$. Similar to Fig.~\ref{figure_focusing}, this figure shows the maximum normalized gain that can be achieved at different distances to the receiver.
The RIS focuses a beam at $(0,0,F)$, for three different $F$, and the signal amplification at other locations $(0,0,z)$ is shown.  The maximum achievable normalized RIS gain is slightly less than the normalized antenna array gain for a SIMO channel in Fig.~\ref{figure_focusing}. The reason for this is the non-zero incident angle $\theta_i$ that leads to a reduction in the integral terms of \eqref{eq:antenna-RIS-gain-exact} due to the varying phase distribution across each RIS element. Other than that, the curves in Fig.~\ref{figure_focusing-RIS} have the same shape as in Fig.~\ref{figure_focusing} for the SIMO channel case, as expected from the fact that the  approximations of the normalized gains that are obtained with continuous matched filtering and optimal continuous RIS phase-shifts are identical and equal to \eqref{eq:G-approx-DF}.
As a result, the DF is the same and the resulting BD can be computed using Theorem~\ref{th:DF}.

\subsection{Beam width in the focal plane}

The beamforming analysis has thus far focused on the limited depth that is unique to near-field beamforming.
We will now consider the transverse BW in the focal plane given by $z=F$, when the RIS is reconfigured for a receiver at $(0,0,F)$. We can compute the 3\,dB BW for an RIS with electrically small elements by again using the Fresnel approximation of $E_r(x,y)$ in \eqref{eq:antenna-RIS-gain-arbitrary-receiver} (as in \eqref{eq:G-approx}) and inject the phase-shift $e^{+\imagunit \frac{2\pi}{\lambda} \left(\frac{x^2}{2F} + \frac{y^2}{2F}+\rho +\sin\left(\theta_i\right)x\right)}$ into the integral  to 
obtain a continuous phase-shift approximation. To obtain the normalized gain variations in the focal plane, we compute the integral over the entire RIS surface for a receiver at $(x_r,y_r,F)$: 
\begin{align} \notag
& \left| e^{-\imagunit \frac{2\pi}{\lambda} (F+\frac{x_r^2+y_r^2}{2F})  } \!\! \int_{- D\sqrt{N/8}}^{D\sqrt{N/8}} \int_{- D\sqrt{N/8}}^{D\sqrt{N/8}}  e^{\imagunit \frac{2\pi}{\lambda}\frac{x_rx+y_ry}{F} } dx \, dy  \right|^2 \\ 
&\times \!\left( \frac{2}{D^2 N} \right)^{\! 2}  = 
\sinc^2 \! \left(\sqrt{\frac{N}{2}}\frac{Dx_r}{\lambda F} \!\right)\sinc^2 \! \left(\sqrt{\frac{N}{2}}\frac{Dy_r}{\lambda F} \! \right) \label{eq:G-approx-BW}
\end{align}
where $\sinc(x)=\sin(\pi x)/(\pi x)$. Notice that $\sinc^2(x)$ is a decreasing function for $x \in [0,1]$ with $\sinc^2(0)=1$ and $\sinc^2(0.443)\approx 0.5$.  Hence, the \emph{$3$\,dB BW} in the focal plane along the $x$-axis is obtained as
\begin{equation} \label{eq:BW}
0.443 \approx \sqrt{\frac{N}{2}}\frac{D|x_r|}{\lambda F} \,\, \rightarrow \,\,  {\rm BW}_{\rm 3 dB}\approx \frac{0.886 \lambda F}{D\sqrt{\frac{N}{2}}}\approx \frac{1.77 F \sqrt{\lambda} }{\sqrt{d_{FA}}}
\end{equation}
where $D\sqrt{N/2}$ is the side length of the RIS.
The 3\,dB BW along the $y$-axis is the same as above. The  formula in \eqref{eq:BW} is the same as for the 3\,dB BW  of a focused aperture in \cite[Eq.~(27)]{Sherman1962a}. Since the distance between the RIS and the focal point is $F$, the \emph{3\,dB angular BW} can be computed  as
\begin{equation} \label{eq:BW-angular}
 {\rm BW}_{\rm 3 dB}^{\rm ang}=2\arctan\left(\frac{ {\rm BW}_{\rm 3 dB}}{2F}\right) \approx 2\arctan\left(\frac{0.886\sqrt{\lambda}}{\sqrt{d_{FA}}}\right) 
\end{equation}
which is the same for any focal distance $F$. However, it decreases with the size of the RIS and the operating frequency. When ${\rm BW}_{\rm 3 dB}$ is  small so that ${\rm BW}_{\rm 3 dB}/(2F)\ll 1$, we can use the approximation $\arctan(x)\approx x$ to get
\begin{equation}
  {\rm BW}_{\rm 3 dB}^{\rm ang} \approx  \frac{ {\rm BW}_{\rm 3 dB}}{F} \approx  \frac{0.886 \lambda }{D\sqrt{\frac{N}{2}}} 
\end{equation}
which is almost the same as the 3\,dB angular BW computed for the far-field beampattern of a uniform linear array with $\sqrt{N}$ elements and inter-element distance $\frac{D}{\sqrt{2}}$ \cite[Eq.~(2.100)]{van2004optimum}.

Since a far-field transmitter was assumed in this section, the phase distribution of the incident field is almost constant over each RIS element and can therefore be compensated for. 
Hence, we can use the same formulas for the DF and BW independently of the incident angle.
Moreover, we note that the BW expressions are the same for the SIMO channel since we end up with the same integral in \eqref{eq:G-approx-BW} when we use the continuous phase-shift approximation in that case.

\begin{figure}[t!]
\begin{center}
	\begin{overpic}[trim={5mm 5mm 15mm 12mm},width=0.95\columnwidth,tics=10]{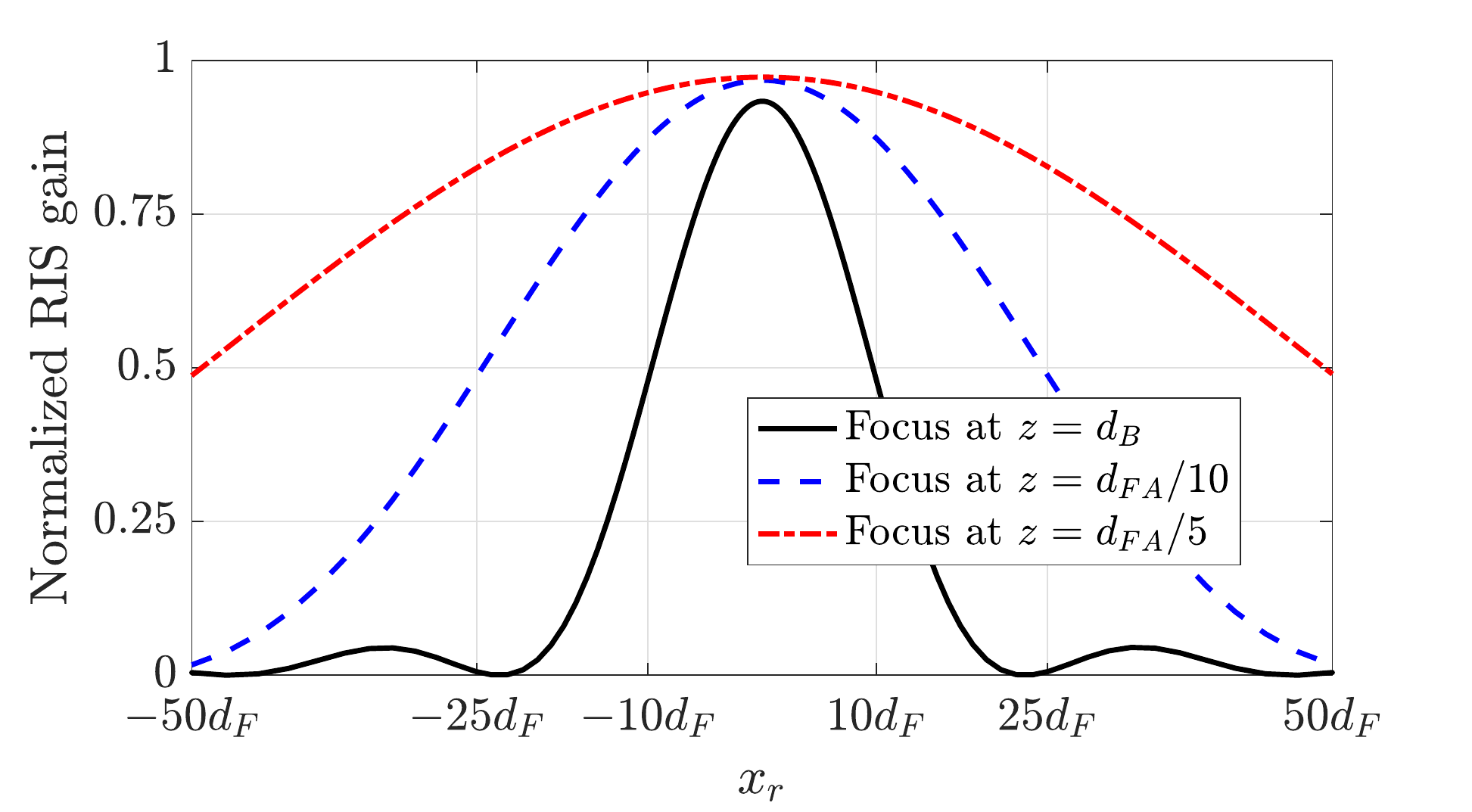}
\end{overpic}  \vspace{-2mm}
\end{center} 
\caption{The BW increases with the distance to the focal point at which the RIS phase-shifts are configured to focus.} \label{figure_BW}   \vspace{-2mm}
\end{figure}

In Fig.~\ref{figure_BW}, we revisit the scenario from Fig.~\ref{figure_focusing-RIS} and plot the normalized RIS gain versus the $x$ coordinate of the receiver, $x_r$.
We consider three focal points $(0,0,F)$ for the RIS beamforming: i)  $F=d_B=400d_F$; ii) $F=d_{FA}/10=1000d_F$; and iii) $F=d_{FA}/5=2000d_F$.
In each case, the receiver is located at $(x_r,0,F)$  in the corresponding focal plane. 
The 3\,dB BW for the case of near-field focusing on $d_B=400d_F$ is roughly $20d_F$ according to \eqref{eq:BW} and it is verified by Fig.~\ref{figure_BW}. When the receiver is located at $x_r=\pm 10d_F$, the RIS gain reduces to half of the value compared to $x_r=0$. When the RIS is configured to focus on $d_{FA}/10=1000d_F$, the BW becomes $50d_F$. The BW has been increased by a factor $1000/400=2.5$ compared to the previous case since it is proportional to the focal distance, $F$, according to \eqref{eq:BW}. When the focal point is doubled to $d_{FA}/5$, the BW is also doubled to $100 d_F$.

\begin{figure}[t!]
        \centering
        \begin{subfigure}[b]{\columnwidth} \centering 
	\begin{overpic}[trim={5mm 5mm 15mm 15mm},width=\columnwidth,tics=10]{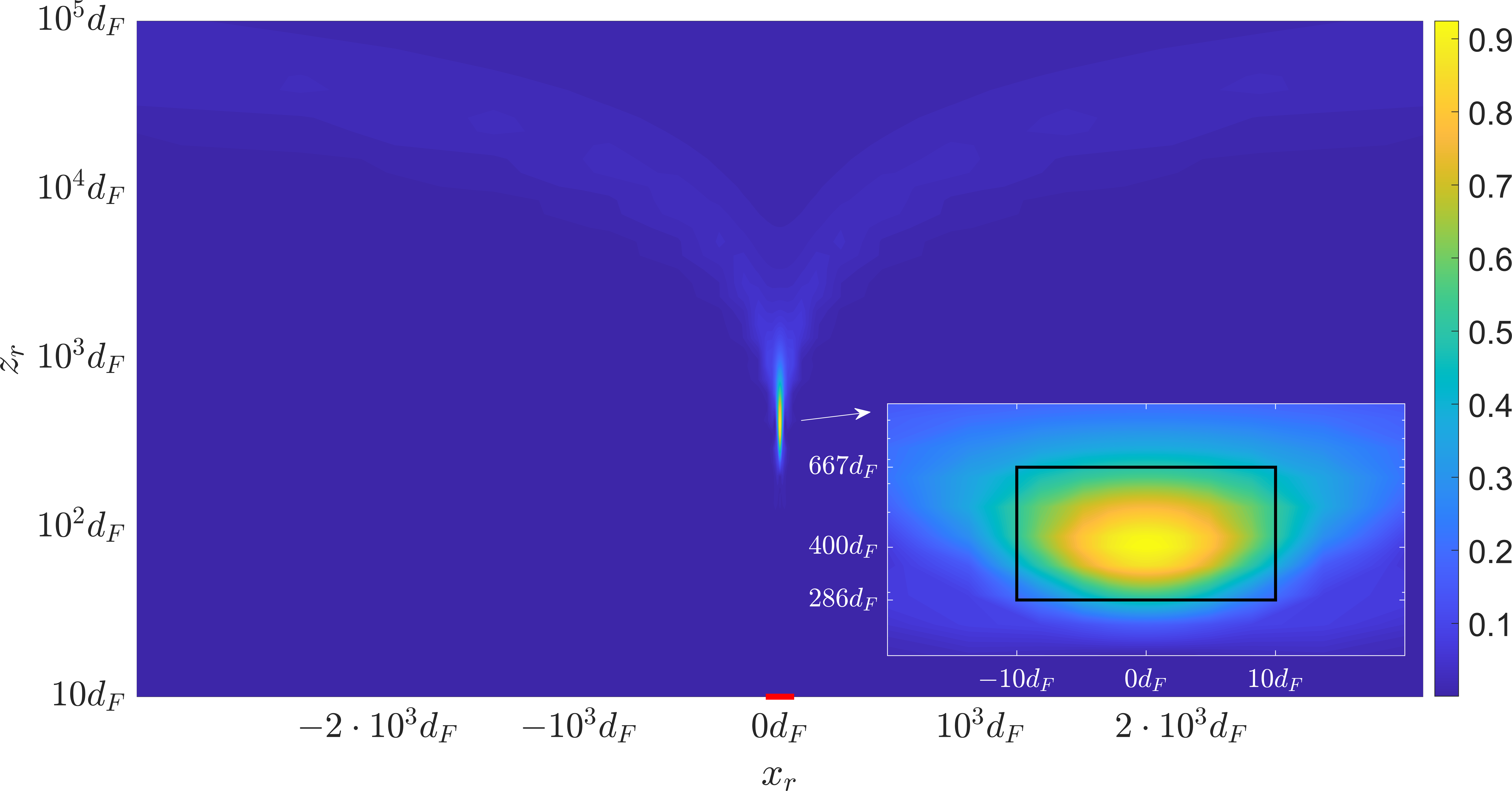}
\end{overpic}   \vspace{-3mm}
                \caption{The RIS focuses a near-field beam on $(0,0,d_B=400d_F)$.} 
                \label{figure_heatmap-focusing-400dF}
        \end{subfigure}\\
        \begin{subfigure}[b]{\columnwidth} \centering  \vspace{+6mm}
	\begin{overpic}[trim={5mm 5mm 15mm 15mm},width=\columnwidth,tics=10]{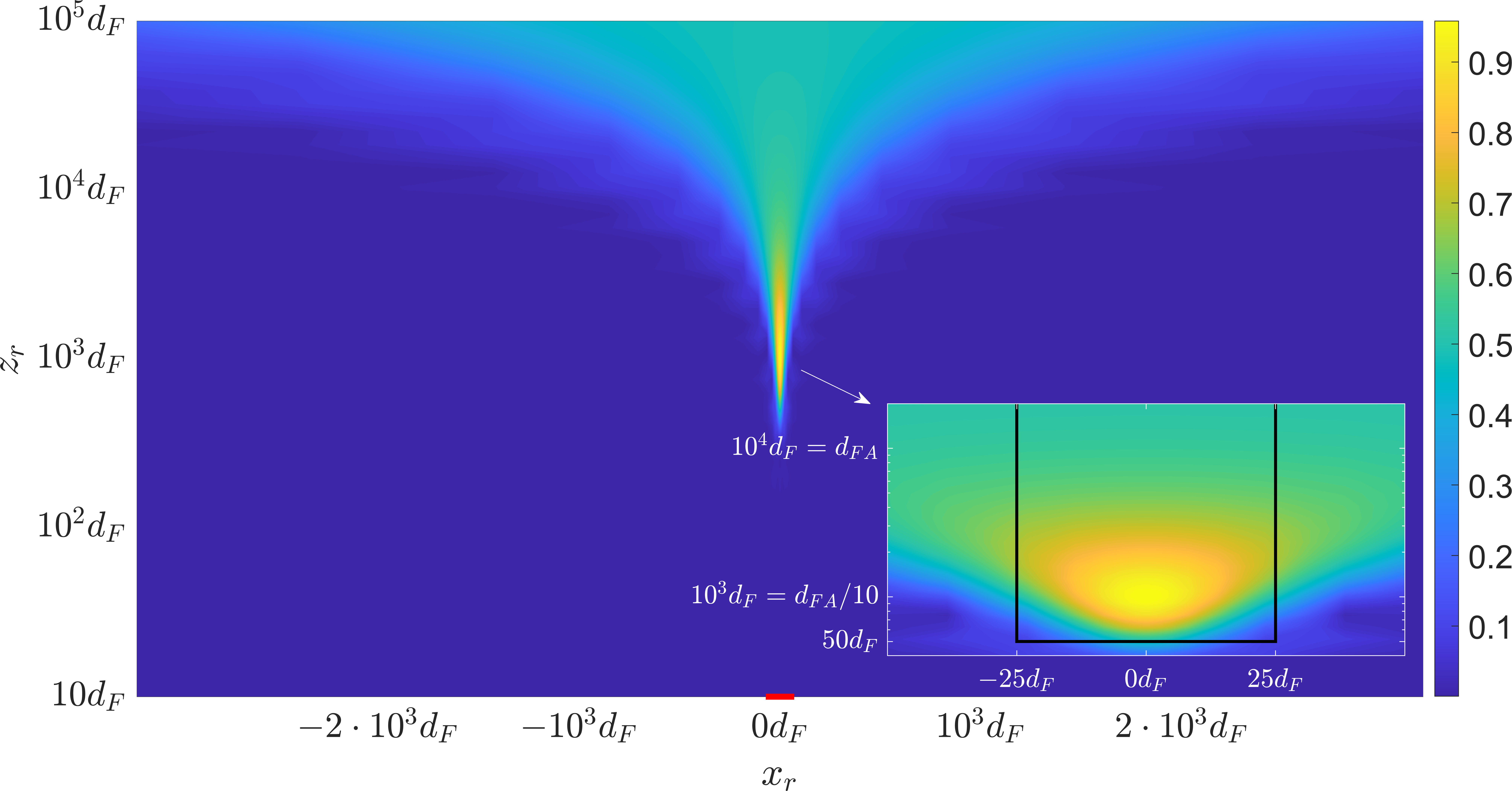}
\end{overpic}   \vspace{-3mm}
                \caption{The RIS focuses a beam on $(0,0,d_{FA}/10=1000d_F)$.} 
                \label{figure_heatmap-focusing-1000dF}
        \end{subfigure} 
        \begin{subfigure}[b]{\columnwidth} \centering  \vspace{+6mm}
	\begin{overpic}[trim={5mm 5mm 15mm 15mm},width=\columnwidth,tics=10]{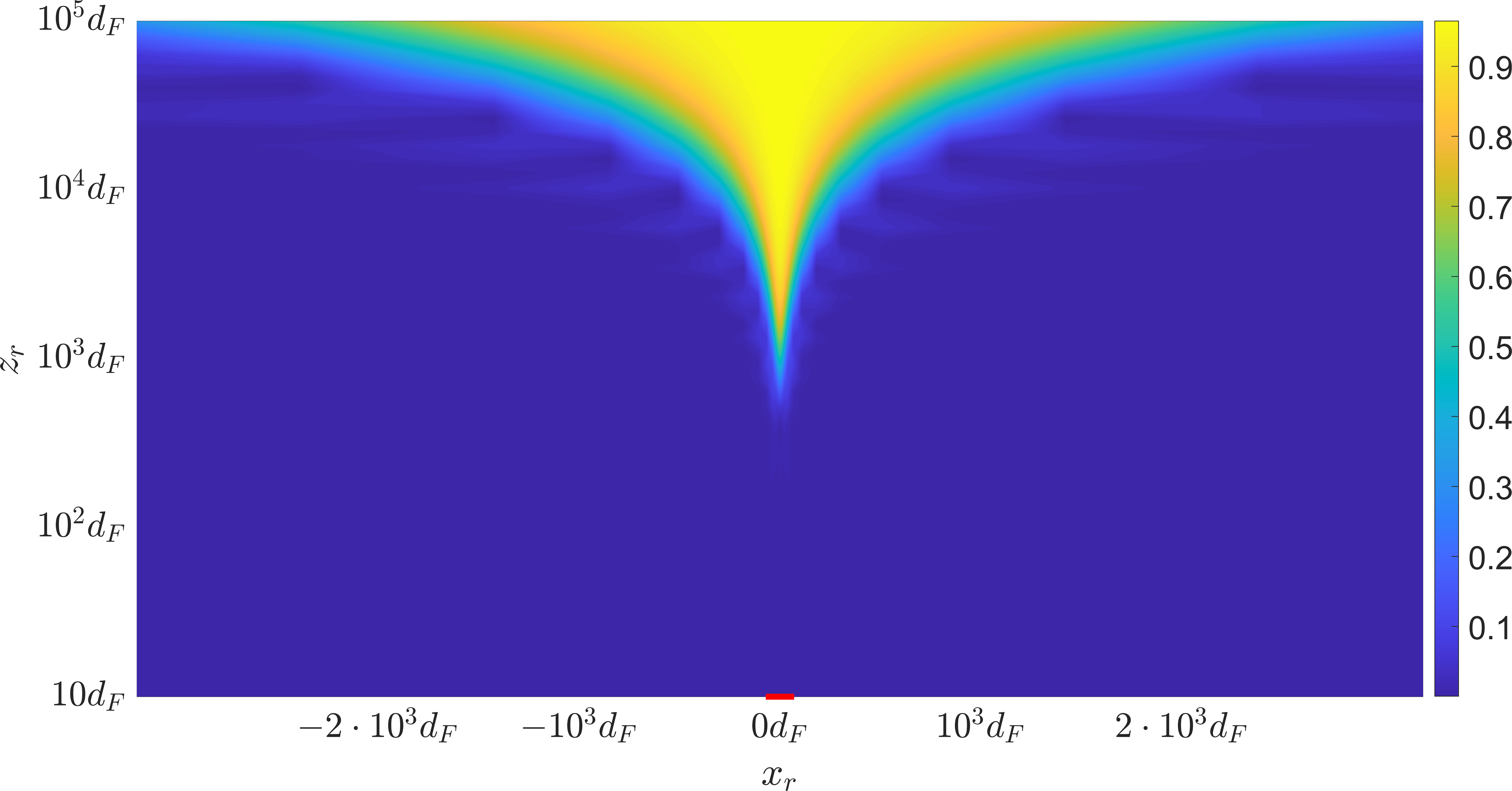}
\end{overpic}  \vspace{-3mm}
                \caption{The RIS focuses a far-field beam on $(0,0,\infty)$.} 
                \label{figure_heatmap-focusing-farfield}
        \end{subfigure} 
        \caption{The heat map of the normalized RIS gain obtained at different receiver locations $(x_r,0,z_r)$ when the RIS is configured to beamform towards three different focal points.} \vspace{-4mm}
        \label{figure_heatmap-focusing}
\end{figure}

Fig.~\ref{figure_heatmap-focusing} shows heat maps of the RIS gains  for three different RIS configurations. The RIS location is shown as a red line at the bottom of the figures. The receiver is located at $(x_r,0,z_r)$ in the $xz$-plane and its coordinates are varied to obtain the heat maps of the RIS gains. In Fig.~\ref{figure_heatmap-focusing-400dF}, the RIS phase-shifts are configured to focus on $(0,0,d_B)$ with $d_B=400d_F$ and in the zoomed plot, the black rectangle specifies the 3\,dB BW in the focal plane along the $x_r$-axis and the 3\,dB BD along the $z_r$-axis. The BW is roughly $20d_F$, which is very narrow. As mentioned before, the DF interval is  $[286 d_F, 667 d_F]$ with the corresponding BD of $381d_F$. For this case, the RIS gain is helpful only in a small region around the focal point.

In Fig.~\ref{figure_heatmap-focusing-1000dF}, the RIS is focused on $(0,0,d_{FA}/10=1000d_F)$, which is beyond the Bj{\"o}rnson distance $d_B$. For this case, the BD is infinite and starts from $z_r=50d_F$. On the other hand, the BW is still finite and it is roughly $50d_F$. Finally, Fig.~\ref{figure_heatmap-focusing-farfield} shows the RIS gains where the RIS is configured to focus on $(0,0,\infty)$. For far-field focusing, both BD and ${\rm BW}_{\rm 3 dB}$ become infinite.  Nevertheless, the angular BW is the same for all the three heat maps since ${\rm BW}_{\rm 3 dB}^{\rm ang}$ in \eqref{eq:BW-angular} is independent of the focal point. The reason that the angular BWs seem to grow in the figures is the logarithmic scale for the $z_r$ axis.

\section{Conclusion}

The near-field and Fresnel region have unequivocal definitions from a transmit antenna perspective, but their implications for communication design and performance are different. This paper is a primer for those analyzing near-field beamforming for antenna arrays and RIS, which is an emerging paradigm where the individual array elements are in the far-field but not the array as a whole.
Conventional far-field beamforming design can be used with at most $3$\,dB loss for propagation distances $z$ larger than $d_{FA}/10$, where $d_{FA}$ is the Fraunhofer array distance. For $z < d_{FA}/10$, the spherical wavefronts must be considered in the beamforming design and the resulting near-field beam has finite depth. 
With optimized beamforming, the maximum antenna array gain can be achieved for  $z \geq d_B$, where $d_B$ is the Bj\"ornson distance and it holds that $d_{FA}\gg d_B$ for large arrays. The angular beam width is the same at all distances but the physical width is small in the near-field. This allows for spatial multiplexing over line-of-sight channels, which is considered in \cite{Dardari2020a,Torres2020a}.

\bibliographystyle{IEEEtran}

\bibliography{IEEEabrv,refs}

\end{document}